\documentclass[aps,prl,twocolumn,showpacs,superscriptaddress]{revtex4}
\usepackage{amsmath}
\usepackage{amssymb}
\usepackage{color}
\usepackage{graphicx}
\usepackage{bm}
\usepackage{epstopdf}

\begin{document}
\newcommand{\beq}{\begin{equation}}
\newcommand{\eeq}{\end{equation}}

\title{The occurrence of a Mott-like gap in single-particle spectra of  electron systems possessing  flat bands }
\author{V.~A. Khodel}
\affiliation{NRC Kurchatov Institute, Moscow, 123182, Russia}
\affiliation{McDonnell Center for the Space Sciences \& Department of Physics,
	Washington University, St.~Louis, MO 63130, USA}

\begin{abstract}
 	An unconventional type of the Mott's insulators where the gap   in the spectrum of single-particle excitations  is associated with repulsive effective interactions between quasiparticles  is shown to exist in strongly correlated electron systems of solids that possess flat bands. The occurrence of this gap is demonstrated to be the consequence
 	 of  violation of particle-hole symmetry, inherent in such systems. The results obtained are
 	   applied to elucidate  the  Fermi arc structure   observed  at temperatures up to $100\, K$ in angle-resolved photoemission spectra
 	    of  the compound  Sr$_2$IrO$_4$, not showing superconductivity down to low $T$.

\end{abstract}
\maketitle

The present article  is devoted to the analysis  of  spectra $\epsilon({\bf p})$ of single-particle excitations of strongly correlated electron systems of solids  within a flat-band  scenario  developed in Refs.  \cite{ks,vol,noz,lee,kats}. In this scenario, non-Fermi-liquid  (NFL) behavior of such systems, experimentally  studied  for   more than 20 years, is  attributed to the occurrence of  flat bands (zero-energy fermions), a $T=0$ dispersionless  portion $\epsilon=0$ of the single-particle spectrum, frequently called  the fermion condensate (FC). Originally, basic  aspects  of theory of  fermion condensation that properly   elucidates  NFL behavior of    strongly correlated Fermi  systems (see e.g. \cite{kss,prb2008,mig100,PRB2012,annals,book,JETPL2015,jetplett2015, vol2015})
   were developed on the base of the  Landau  approach to FL theory where the ground state energy $E$   is treated as a functional of the quasiparticle momentum distribution $n({\bf p})$. A crucial  point  is that the ground-state
  momentum distribution $n_*({\bf p}) $ of  a system with the FC is found with the aid of  variational condition \cite{ks}
  \beq
  \frac{\delta E}{\delta n({\bf p})}-\mu=0 , \quad {\bf p}\in \Omega .
  \label{var}
  \eeq
   Since in  normal states, examined in this article,  the l.h.s. of this condition is nothing but the quasiparticle energy $\epsilon({\bf p})$ measured from the chemical potential $\mu$, this equation implies  the formation of the FC
  in the momentum region ${\bf p}\in \Omega$ where   $n_*({\bf p}) $   changes  continuously  between 0 and 1. In the complementary domain ${\bf p}\notin\Omega$,    associated with quasiparticles, not belonging to the FC,   the distribution $n_*({\bf p})$ coincides with the FL one,  being 1 for the occupied states and 0, otherwise.

 Within theory of fermion condensation [1-5],    the dispersion of the   spectrum of such quasiparticles, called further normal, is evaluated   in terms of  a phenomenological interaction function $f({\bf p},{\bf p}_1)$ with the aid of Landau equation
   \beq
   \frac {\partial \epsilon({\bf p})}{\partial {\bf p}}= \frac{{\bf p}}{M}+\int f({\bf p},{\bf p}_1)\frac {\partial \epsilon({\bf p}_1)}{\partial {\bf p}_1} d{\bf p}_1 , \quad {\bf p}\notin \Omega ,
   \label{laneq}
   \eeq
where $d{\bf p}$ is the volume element in momentum space,
  including the factor $(2\pi)^i$  in the denominator, with $i$, being dimensionality of the problem.

Results of numerous calculations (see e.g. Ref.\cite{prb2008}) demonstrate that  there is   no gap, separating this normal part of the spectrum $\epsilon({\bf p})$ from
  the dispersionless FC  one. The purpose of the present article is to check whether this feature holds,  going
    beyond  the scope of the  existing version of theory of fermion condensation. As we will see,  it does not: if    interactions between the FC and normal quasiparticles are taken into account properly,   a  gap  emerges that separates the FC spectrum  from that of normal quasiparticles.

   Here we address the case $T=0$. Since at finite $T$, the FC dispersion changes linearly with $T$ \cite{noz}, the $T=0$ results obtained   below hold at low $T$, as long as the gap  value $D(0)$   exceeds  the FC width
 $\propto \rho_{FC} T$. Otherwise,  corrections to results obtained within the existing version of theory  are of no interest, being small.

     To gain insight into the problem it is advantageous to employ
    the Belyaev's diagram technique developed
   in his work on theory of Bose liquid \cite{belyaev1}. In doing so we   treat results of  numerical solving  the set of Eqs. (\ref{var}) and (\ref{laneq}) for the momentum distribution $n_*({\bf p})$ and   energy spectrum, denoted further $\epsilon_0({\bf p})$, as an initial  iterate.

    Since   every integration over the FC domain introduces  an additional  small dimensionless factor $\eta=\rho_{FC}/\rho$, the full set of   the diagrams under consideration     can be divided into subsets  vs.  the amount of  the FC lines displayed.   This   situation is   opposite to that in low-density Bose gas where just  occupation numbers of normal quasiparticles, proportional to the difference between the  total density and  condensate one,  are small \cite{belyaev2}.

  The simplest way  to generalize the existing version of theory of fermion condensation  is to straightforwardly  evaluate  the imaginary part  $\Sigma''$ of the mass operator of the normal  quasiparticle and then  calculate  the real part $\Sigma'$   with the aid of the Kramers-Kronig dispersion relation. Therefore
    it is instructive to begin  the analysis  with   remembering basic points of   evaluation   of   $\Sigma$  in FL theory  where formula for  $\Sigma''$  reads \cite{galitskii}:
$$
 \Sigma''({\bf p},\varepsilon)\propto - \int\int |\Gamma^2({\bf p},{\bf p}_1,{\bf p}_2,{\bf p}_3)|\delta( \varepsilon+\epsilon_3
-\epsilon_1-\epsilon_2)$$
\beq
\biggl ( n_3(1-n_1)(1-n_2)-
(1-n_3)n_1n_2\biggr)
d{\bf p}_1 d{\bf p}_2.
\label{gal}
\eeq
Here  $\Gamma$ stands for the scattering amplitude, and $n_k=\theta(-\epsilon_k)$   are $T=0$ quasiparticle occupation numbers,  where  $\epsilon_k=\epsilon({\bf p}_k)$, with $k=1,2,3$ and  ${\bf p}_3={\bf p}_1+{\bf p}_2-{\bf p}$,  are  single-particle energies.

  Strictly speaking, the  exact  $T=0$ formula  for   $\Sigma''$ does contain the product of three spectral functions
\beq
A({\bf p},\varepsilon)\propto  \frac{ |\Sigma''({\bf p},\varepsilon)|}
{[\varepsilon-\epsilon_0({\bf p})-\Sigma'({\bf p},\varepsilon)]^2+ [\Sigma''({\bf p},\varepsilon)]^2}  ,
\label{img}
\eeq
   associated with imaginary parts of respective   quasiparticle Green functions
    $G({\bf p},\varepsilon)=(\varepsilon-\epsilon_0({\bf p})-\Sigma({\bf p},\varepsilon))^{-1}$.
   However,   in conventional Fermi liquids, the
damping $\gamma(\varepsilon)$ of single-particle excitations is quadratic in energy:
\beq
\gamma(\varepsilon)\propto  -\Sigma''(\varepsilon>0)\propto  \varepsilon^2 ,
\label{imsfl}
\eeq
implying that   it  is small  compared with  energy.
 Indeed,  in Eq.(\ref{gal})  integration  virtually occurs over 3 positive energies $\epsilon_1,\epsilon_2$ and $-\epsilon_3$,  confined to the interval  $[0,\varepsilon]$, the number of integrations  reducing to 2   by virtue of the  presence of  $\delta( \varepsilon+\epsilon_3
 -\epsilon_1-\epsilon_2)$ in the integrand.  As a result, each of two remaining integrations   introduces the factor $\varepsilon$ to yield  Eq.(\ref{imsfl}) and justify the  replacement $A({\bf p},\varepsilon)\to  \delta(\varepsilon-\epsilon({\bf p}))$.

  Once the imaginary part $\Sigma''$ of the mass operator changes continuously through  the Fermi surface, so does its  real part $\Sigma'({\bf p},\varepsilon)$ as well. In this case,   the single-particle spectrum  $\epsilon({\bf p})$, evaluated
   from   standard equation
   \beq
   \epsilon({\bf p})=\epsilon_0({\bf p})+\Sigma'({\bf p},\epsilon({\bf p})),
   \label{ssp}
   \eeq
   with the  bare spectrum $\epsilon_0({\bf p})$, turns out to be gapless.

   However,  in  Fermi systems with flat bands, Eq.(\ref{imsfl}) {\it   fails}, since
   in  calculations of  Eq.(\ref{gal}) two  energies  associated with  FC quasiparticles  {\it identically   vanish}, so that the number of energy integrations reduces from 3 to 1,  and  consequently,   the factor  $\varepsilon^2$, identifying conventional Fermi liquids,   disappears.   As a result, the damping $\gamma(\varepsilon)$  turns out to be {\it energy independent}, and therefore
   \beq
    \Sigma''(\varepsilon\to 0)\propto -\frac {\varepsilon }{|\varepsilon |}
    \label{nflis}
   \eeq
  that rules out  the conjecture $A({\bf p},\varepsilon)\to  \delta(\varepsilon-\epsilon({\bf p}))$.

  Nevertheless, the  result  (\ref{nflis})  itself remains unchanged \cite{PRB2012}. Indeed,  by virtue of the dispersionless character of the FC spectrum, this subsystem behaves as a set of impurities. Therefore  in the  amplitude of  scattering  of normal quasiparticles by the FC, there is  a pure elastic term. This circumstance   straightforwardly leads to 	Eq.(\ref{nflis}).
  	
  	In Fermi gas with impurities, the role of the real part $\Sigma'$ of the mass operator $\Sigma$ reduces to a slight renormalization of the chemical potential $\mu$. Contrariwise,  in Fermi systems with flat bands, $\Sigma'(\varepsilon)$  acquires a logarithmically divergent   term
  	\beq
  	\Sigma'(\varepsilon\to 0)\propto -\frac{ \varepsilon}{ |\varepsilon|}\ln |\varepsilon| ,\
  	\label{nflrs}
  	\eeq
  	 due to {\it violation of  particle-hole symmetry}, inherent in these systems, (see below). This implies the   occurrence of the gap in the single-particle spectrum, verified  by inserting Eq.(\ref{nflrs}) into Eq.(\ref{ssp}), a basic result of the
 analysis performed in the present article.

\begin{figure} [! ht]
    	\begin{center}
    		\includegraphics [width=0.3\textwidth]{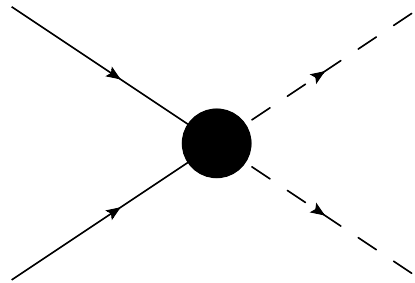}
    	\end{center}
    	\caption{ The graphical representation for   the transfer of two normal quasiparticles  (solid lines) to the FC
    		(dotted lines).     		}\label{fig1}
    \end{figure}

 Let us now turn to a more detailed analysis  of  second-order FC contributions to the imaginary part $\Sigma''$ of the mass operator $\Sigma$, coming  from  diagrams that contain two FC lines. (The total contribution  of diagrams with the single FC line   was shown  long   ago    not to provide    the gap in the spectrum   $\epsilon({\bf p})$  \cite{kss,kz}).
 The secon-order part of $\Sigma''$ is    found with the aid of a modified  formula (\ref{gal})  where  two functions $n({\bf p})$ are replaced by $n_*({\bf p})$. There are several options to do that. However,  violation of  particle-hole symmetry, discussed above, occurs only in a  diagram displayed in Fig.~\ref{fig1} where  normal  quasiparticles, depicted by solid lines, convert to the FC quasiparticles, drawn by dashed ones. This diagram is   reminiscent   of that,  relevant to   inhomogeneous Larkin-Ovchinnikov-Fulde-Ferrell (LOFF) pairing   with  certain  total momentum ${\bf P}\neq 0$ \cite{larkin,fulde}. However,  in the case under consideration, where LOFF pairing is supposed to be forbidden by virtue of the repulsive character of the interaction between quasiparticles in the Cooper channel,  integration over all  accessible momenta ${\bf P}$ is carried out that provides the restoration of  homogeneity of the ground state.
      The  explicit expression for $\Sigma''({\bf p}\notin \Omega,\varepsilon)$ is obtained from  Eq.(\ref{gal}) with  replacements:  i)$\epsilon_1=\epsilon_2\to 0$, $ n({\bf p}_1)\to n_*({\bf p}_1), n({\bf p}_2)\to n_*({\bf p}_2)$,
       ii)$\delta(\varepsilon+\epsilon({\bf p}_3))\to A({\bf p}_3,-\varepsilon)$, and iii)
 $A({\bf p}_1\in\Omega,\varepsilon_1)\to \delta(\varepsilon_1) $, $ A({\bf p}_2\in\Omega,\varepsilon_2)\to \delta(\varepsilon_2)$, (the latter replacement holds  provided fourth-order FC
   contributions are neglected).

            As a result, we find
        $$
        |\Sigma''({\bf p},\varepsilon>0)|= \int_C |\Gamma^2({\bf p},{\bf p}_2,{\bf P}-{\bf p},{\bf P}-{\bf p}_2)|$$
        \beq
        (1-n_*({\bf p}_2))(1-n_*({\bf P}-{\bf p}_2)) A({\bf P}- {\bf p},-\varepsilon) d{\bf P} d{\bf p}_2,
        \label{imsp}
        \eeq
        while
        $$
         |\Sigma''({\bf p},\varepsilon<0)|=
        \int_C |\Gamma^2({\bf p},{\bf p}_2,{\bf P}-{\bf p},{\bf P}-{\bf p}_2)|$$
        \beq
        n_*({\bf p}_2)
        n_*({\bf P}-{\bf p}_2)A({\bf P}-{\bf p},-\varepsilon) d{\bf P} d{\bf p}_2.
        \label{imsn}
        \eeq
    From  aforesaid  we infer   that    the scattering amplitude $\Gamma$, entering this expression,
        is  evaluated at $\eta=0$.
             The   domain  $C$ of  integration over  momenta ${\bf p}_2$  and ${\bf P}$ is determined   by 	 conditions
             \beq
             {\bf p}_2\in \Omega,\quad   {\bf P}-{\bf p}_2\in \Omega .
             \label{fcbp}
             \eeq

  As seen, the integration over ${\bf p}_2$ is separated from that over ${\bf P}$, so that upon inserting the explicit form of the spectral function $A$ we arrive at
           \beq
           | \Sigma''({\bf p},\varepsilon)|=
           \int\frac{ K({\bf p},{\bf s},\varepsilon)|\Sigma''({\bf s},-\varepsilon)|d {\bf s}  } {(e({\bf s},\varepsilon)+\epsilon_0({\bf s }))^2+( \Sigma''({\bf s},-\varepsilon))^2} ,
           \label{imsin}
           \eeq
     where      ${\bf s}={\bf P}-{\bf p}$ and
     \beq
     e({\bf s},\varepsilon) =\varepsilon+\Sigma'({\bf s},-\varepsilon) ,
     \label{defe}
     \eeq
      while  the normal  component $\epsilon_0({\bf p})$  of the  single-particle spectrum is found from
      Eq.(\ref{laneq}). The function  $K(\varepsilon)$ is defined as
       $K(\varepsilon>0)=K^+$, and   $K(\varepsilon<0)=K^-$, with
   \begin{eqnarray}
   K^+&=&   \int_C
   |\Gamma^2({\bf p},{\bf p}_2,{\bf P})| (1- n_*({\bf p}_2))
   (1- n_*({\bf P}-{\bf p}_2))d{\bf p}_2 ,\nonumber\\
   K^-&=&   \int_C
   |\Gamma^2({\bf p},{\bf p}_2,{\bf P})|  n_*({\bf p}_2)
   n_*({\bf P}-{\bf p}_2)d{\bf p}_2 .
   \label{kpm}
   \end{eqnarray}
  Evidently, both the functions $K^{\pm}$ change linearly with the FC density $\eta$.

   The presence  of two different expressions, containing  the FC momentum distribution $n_*({\bf p})$ in these formulas does  ensure {\it violation of particle-hole symmetry}, since the FC  distribution is not invariant with respect to  the replacement  $n_*({\bf p})\to 1-n_*({\bf p})$,  and then
        $  | \Sigma''({\bf p},\varepsilon\to 0^+)|\neq  | \Sigma''({\bf p},\varepsilon\to 0^-)|$.
       We will employ this fact in derivation of Eq.(\ref{nflrs}).

          In what follows we focus on
          2D electron liquid, placed in an external field of the quadratic lattice; such a  situation is relevant to  cuprates, the most extensively studied family of high-$T_c$ superconductors. In this case,
          the  FC domain consists of four small spots \cite{zkc}, adjacent to saddle points $(0,\pm \pi), (\pm \pi,0)$, associated with van Hove points (VHPs) where the density of states  diverges  (see Fig.~2).
          The association  between the FC spots and VHPs stems from observation  that the onset of  fermion condensation is triggered by  violation  of  the {\it necessary stability condition}  for the Landau state  (for detail, see e.g. Ref.\cite{physrep}), occurring just beyond  critical points where the density of states diverges.

          First, let  momenta
          ${\bf p}_2$  and ${\bf P}-{\bf p}_2$  be related to    opposite FC spots.  Total momentum ${\bf P}$ is then close to 0 that resembles  the case of Cooper pairing.
          Boundaries of the  integration region $C$
          are found with the aid of  Eqs. (\ref{fcbp}).  In the case where  momentum ${\bf p}_2$ is related  e.g. to the  FC spot, situated near the saddle point $(0,\pi)$, while  momentum  ${\bf P}-{\bf p}_2$,  to the  FC spot,  located close to the opposite saddle point $(0,-\pi)$,  one obtains
          \begin{eqnarray}
          -\frac{L_x}{2}&\leq&(P_x-p_{2x})\leq \frac{L_x}{2},
          -\pi\leq(P_y- p_{2y})\leq -\pi +\frac{L_y}{2},\nonumber\\
          -\frac{L_x}{2}&\leq& p_{2x}\leq \frac{L_x}{2},  \quad  \pi-\frac{L_y}{2}\leq p_{2y}\leq \pi ,
          \label{r11}
          \end{eqnarray}
          so  that
          \beq
          -L_x\leq P_x\leq L_x ,\quad  -L_y/2\leq P_y\leq L_y/2,
          \label{rf}
          \eeq
          where quantities $L_x,L_y\simeq \eta^{1/2}$  determine the FC range.
    	\begin{figure}
    		\includegraphics[width=0.8\linewidth,height=0.8\linewidth]{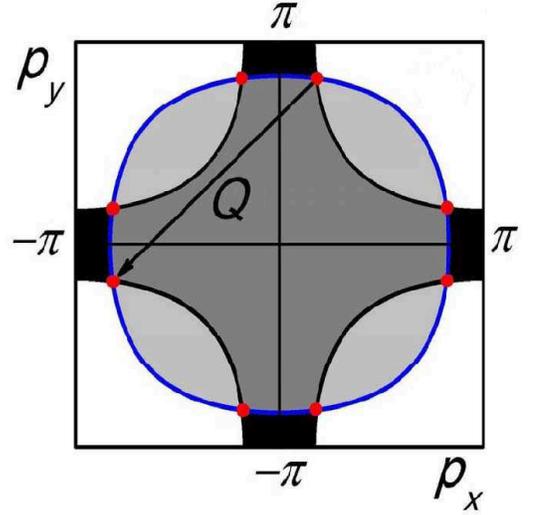}
    \caption{Fermi line (black) and its counterpart (blue) for the bare
tight-binding spectrum
$\epsilon^0_{\bf p} = -2\,t_0\,(\cos p_x
+ \cos p_y)+4\,t_1\,\cos p_x\cos p_y$,
with $t_1/t_0=0.45$. The FC regions \cite{zkc} are colored in black. }
    	\end{figure}
    	
         The implementation  of the  mean value theorem (MVT) for integrals, applicable due to positivity of all functions, standing in the integrand, allows one to recast   Eq.(\ref{imsin}) in the form
          \beq
          | \Sigma''({\bf p},\varepsilon)|= K_{av}({\bf p},\varepsilon)
          \int\frac{|\Sigma''({\bf s},-\varepsilon)|d {\bf s}  } {(e({\bf s},\varepsilon)+\epsilon_0({\bf s }))^2+( \Sigma''({\bf s},-\varepsilon))^2}
          \label{imsin11}
          \eeq
   where $ K_{av}({\bf p},\varepsilon)\propto \eta$
          is an averaged value of the function $K({\bf p},{\bf P},\varepsilon) $  in the integration region $C$,    controlled by Eq.(\ref{fcbp}).

    Inequalities (\ref{rf}) imply that the distance between  vectors ${\bf p}$ and ${\bf s}$ is small that allows one to replace the function $\Sigma({\bf s})$, standing in the denominator of the integrand (\ref{imsin11}), by $\Sigma({\bf p})$.  Furthermore, in the region near the Fermi surface where  $|\varepsilon|,|\epsilon_0({\bf p})|\leq v_FP\simeq \epsilon^0_F\eta$, both   the term $\Sigma'$ and $\varepsilon$ can be neglected, (see below).
     With this simplifications,  Eq.(\ref{imsin}) becomes
           \beq
       | \Sigma''({\bf p},\varepsilon)|= K_{av}({\bf p},\varepsilon)
       \int\frac{|\Sigma''({\bf p},-\varepsilon)|d {\bf s}  } {\epsilon_0^2({\bf s })+( \Sigma''({\bf p},-\varepsilon))^2} .
       \label{imsin1}
       \eeq

       The two-dimensional  integration  in this equation,
       whose limits are specified by Eq.(\ref{rf}),  is performed    with the aid of formula
       $d{\bf s}=dn dt=d\epsilon_0 dt/|{\bf v}_0| $, where
       $n$ and $t$ are  normal and transversal components of the vector ${\bf s}$,   while ${\bf v}_0({\bf s})=\nabla \epsilon_0({\bf s})$ where ${\bf s}\simeq {\bf p}\notin \Omega$.  The repeated application of
              the MVT then  yields
  \begin{eqnarray}
  \Sigma''({\bf p},\varepsilon>0)&=& -\pi l \zeta({\bf p}) K_{av}^+({\bf p}) ,\nonumber\\
 \Sigma''({\bf p},\varepsilon<0)&=& \pi l \zeta({\bf p}) K_{av}^-({\bf p}),
  \label{imsin3}
  \end{eqnarray}
where $l\simeq L\propto \eta^{1/2}$ is the length of the interval of  integration over $t$, and $\zeta({\bf p})$, an averaged value of the  function $1/v_0({\bf s})$.

Evidently, the result obtained is in agreement with  Eq.(\ref{nflis}), discussed above, justifying that in systems with flat bands, the absolute value of the imaginary part of the mass operator of a normal quasiparticle experiences a {\it  discontinuity } at the Fermi surface. Its magnitude, being  of order  $\eta^{3/2}$, consists of  the factor $\eta$, coming  from  integration over ${\bf p}_2$,  and an additional factor
 $\eta^{1/2}$ associated with the limits of integration over $t$.  Similar results  are obtained in case the FC momenta belong to the same FC spot, since then  total momenta ${\bf P}$ turn out to be close to $2\pi/a$, and  integration is performed over a small region of  momenta ${\bf P}'=2\pi/a-{\bf P}$. At the same time,  contributions from  neighbor FC spots are verified to be suppressed.

 Having at hand these results,
the real part $\Sigma'$ of the mass operator is then found on the  base of the Kramers-Kronig dispersion relation
\beq
\Sigma'(\varepsilon)=\frac{1}{\pi}P \int\limits_{-\infty}^\infty \frac {\Sigma''(\varepsilon')sign (\varepsilon')}{\varepsilon'-\varepsilon} d\varepsilon' .
\label{dr}
\eeq
 In conventional Fermi liquids, the particle-hole symmetry holds, implying that $\Sigma''(\varepsilon\to 0^+)=-\Sigma''(\varepsilon\to 0^-)$, and  therefore the integral (\ref{dr})  identically vanishes. True, at large distances from the Fermi surface, this symmetry is somehow violated. However, this violation  leads only to  a renormalization of the chemical potential $\mu$.
     Contrariwise,  according to Eqs.(\ref{kpm}) and (\ref{imsin3}), in systems with flat bands,  violation of the particle-hole symmetry occurs just at the Fermi surface.
This makes the difference. Indeed,
upon inserting Eq.(\ref{imsin3}) into Eq.(\ref{dr}) and simple manipulations we are led to
 \begin{eqnarray}
 \Sigma'({\bf p},\varepsilon\to 0^+) & =&-\lambda \eta^{3/2} \zeta({\bf p})\ln|\varepsilon| , \nonumber\\
 \Sigma'({\bf p},\varepsilon\to 0^-) & =&\lambda\eta^{3/2}  \zeta({\bf p})\ln|\varepsilon| .
 \label{resigf}
 \end{eqnarray}
  In writing Eq.(\ref{resigf})    all  numerical factors, independent of $\eta$,     are absorbed into the effective coupling constant $\lambda\propto K_{av}^-({\bf p})-K^+_{av}({\bf p})$, whose sign is supposed to  be positive to  avoid contradictions  with the requirement $\partial \Sigma'(\varepsilon)/\partial\varepsilon<0$.

   With the results obtained, approximations made above are easily verified.  Indeed, according to Eq.(\ref{resigf}),  one has $\Sigma'({\bf s})\propto \eta^{3/2}$. At the same time,   $|\epsilon_0({\bf s})|\simeq Pv_F\propto \eta $, so that   the contribution  from $\Sigma'$ to Eq.(\ref{imsin1}) can be freely neglected. The same is valid for the  term $\varepsilon$, as long as $|\varepsilon|<  Pv_F$.

 Upon inserting Eq.(\ref{resigf}) into Eq.(\ref{ssp}) where $\varepsilon$ is replaced by the true single-particle energy, denoted further by $E$, we are led to
    \begin{eqnarray}
    E({\bf p}) + \lambda\eta^{3/2}\zeta({\bf p})\ln E({\bf p}) &=&\epsilon_0({\bf p}), \quad E>0 , \nonumber\\
   E({\bf p}) -\lambda\eta^{3/2}\zeta({\bf p})\ln |E({\bf p})| &=&\epsilon_0({\bf p}), \quad E<0 .
    \label{eqgap}
    \end{eqnarray}
 Upon setting $\epsilon_0({\bf p})=0$,
     two nontrivial solutions of  Eq.(\ref{eqgap}) are found:
     \beq
     E({\bf p})\simeq\pm\lambda\eta^{3/2}\zeta({\bf p})\ln  \left(1/\eta\right).
     \label{mgap}
     \eeq
We emphasize that the occurrence of the  gap  (\ref{mgap}) in the  single-particle spectrum is entailed by the divergence (\ref{resigf}) of the real part of the mass operator, (cf.  situation in BCS theory where gap solutions $E({\bf p})=\pm \Delta({\bf p})$ owe their existence  to the pole singularity of the Cooper  mass operator $\Sigma'(p,\varepsilon)=\Delta^2/(\varepsilon+\epsilon_0(p))$).    Since the gap (\ref{mgap}) emerges in the normal state,  it can be viewed as a unconventional Mott's gap  in the spectrum of single-particle excitations of  systems possessing  flat bands.

Noteworthy, in deriving these  results we applied the same perturbation-theory strategy as J.\ Kondo in his seminal work on the problem of electron scattering  by  magnetic impurities in metals. Curiously,  his result also contains the logarithmic term $\ln |\varepsilon|$, however,  in contrast to Eq. (\ref{resigf}), the Kondo  correction enters the imaginary part of the electron  mass operator, rather than the real one.  Summation of higher orders of  the Belyaev-like   expansion employed here  is beyond  the scope of the present article.  Nevertheless, we hope that
similarly to the situation with the Kondo effect,
 such a  summation reduces only  to the renormalization of input parameters.

The  gap (\ref{mgap})  has the specific angular dependence  associated with
  the  factor $\zeta ({\bf n})$,  (${\bf n}={\bf p}/p$). Outside the FC regions,   this quantity   differs little from
  $1/v_0({\bf n})$. Therefore by virtue of  vanishing of   $v_0({\bf p}\in \Omega)$  due to the dispersionless character of the FC spectrum, the gap magnitude rapidly grows toward  the saddle points  $(0,\pm \pi), (\pm \pi,0)$ that results in a specific Fermi arc structure (FAS) of the angle-resolved photoemission spectrum,
    breaking up of the Fermi surface into disconnected  segments. Usually this structure   exhibits itself  in the ARPES data  on high-$T_c$ superconductors where  it
      is conventionally   attributed to  the occurrence of   preformed pairs \cite{norman}. However, recently the FAS was uncovered  in measurements of photoemission spectra of  a 2D metal Sr$_2$IrO$_4$ that shows  no superconductivity down to low $T$; nevertheless, the FAS persists up to 100\, K \cite{kim,baumberger,denlinger} that rules out the conventional scenario \cite{norman}.

It is instructive to address  the situation where the  Mott's gap is sufficiently large to provide
profound  suppression of the  conductivity $\sigma(T)\propto e^{-D/T}$. In this case,  the  electron system with the flat band behaves as  a  strongly correlated system of   {\it neutral fermions}, with the magnetic moment $\mu_B\propto e/m_e$. Its thermodynamic properties, associated with the flat band,  remain the same as   in the situation without the Mott's gap.

  Calculations, carried out  above, can also be performed  in the case of homogeneous matter to yield
  \beq
  E( p)\simeq\pm\lambda\eta\frac{\ln  \left(1/\eta\right)}{v_F} .
  \label{mgaph}
  \eeq
   With respect to Eq.(\ref{mgap}), the corresponding value of the Mott's gap is somewhat  enhanced
  by virtue of different kinematic   restrictions in comparison to Eq.(\ref{rf}).
  The presence of the significant Mott's gap may affect  properties of  dense quark matter where the FC presumably resides \cite{prb2008}. In the  scenario, discussed in the present article, opening the Mott's gap  results in the  profound suppression of neutrino cooling of  hybrid compact stars with a sharp hadron-quark interface.  In connection with this idea,  it is not improbable that this suppression is relevant to   recent observations of a central compact object in the supernova remnant   HESSJ1731-347, being the hottest isolated neutron star, in spite of its venerable age \cite{ksp,okk}.

I thank A. Chubukov,  V. Shaginyan, G. Volovik and M. Zverev for valuable discussions. I am also grateful
   to D. Page for discussing  key points of the modern scenario for neutrino cooling  of neutron stars.

This work was partially supported by the Russian Foundation
 for Basic Researches, (project nos. 14-02-00107-a and 15-02-06261) and by the Council of the President of the Russian Federation for Support of  Young Scientists and Leading Scientific  Schools,
 (project no. NMSh-932-2014.2).
 I also thank the McDonnel Center for the Space Sciences
  for  support.

\end{document}